# Dual-Phase High-Entropy Ultra-High Temperature Ceramics

Mingde Qin [a], Joshua Gild [a], Chongze Hu [a], Haoren Wang [b], Md Shafkat Bin Hoque [c], Jeffrey L. Braun [c], Tyler J. Harrington [a, b], Patrick E. Hopkins [c, d, e], Kenneth S. Vecchio [a, b], Jian Luo [a, b, *]

[a] Program of Materials Science and Engineering, University of California, San Diego, La Jolla, CA, 92093, USA
[b] Department of NanoEngineering, University of California, San Diego, La Jolla, CA, 92093, USA
[c] Department of Mechanical and Aerospace Engineering, University of Virginia, Charlottesville, VA, 22904, USA
[d] Department of Materials Science and Engineering, University of Virginia, Charlottesville, VA, 22904, USA
[e] Department of Physics, University of Virginia, Charlottesville, VA, 22904, USA

## Abstract

A series of dual-phase high-entropy ultrahigh temperature ceramics (DPHE-UHTCs) are fabricated starting from $N$ binary borides and (5-$N$) binary carbides powders. >~99% relative densities have been achieved with virtually no native oxides. These DPHE-UHTCs consist of a hexagonal high-entropy boride (HEB) phase and a cubic high-entropy carbide (HEC) phase. A thermodynamic relation that governs the compositions of the HEB and HEC phases in equilibrium is discovered and a thermodynamic model is proposed. These DPHE-UHTCs exhibit tunable grain size, Vickers microhardness, Young' and shear moduli, and thermal conductivity. The DPHE-UHTCs have higher hardness than the weighted linear average of the two single-phase HEB and HEC, which are already harder than the rule-of-mixture averages of individual binary borides and carbides. This study extends the state of the art by introducing dual-phase high-entropy ceramics (DPHECs), which provide a new platform to tailor various properties via changing the phase fraction and microstructure.

---

## 1. Introduction

The introduction of high-entropy alloys (HEAs) by Cantor et al. [1] and Yeh et al. [2] in 2004 has attracted significant research interest in the physical metallurgy community. HEAs can be considered as a subset of compositionally-complex alloys (CCAs) or multi-principal element alloys (MPEAs) [3]. Specifically, a HEA typically has five or more constituent elements of equal or nearly-equal molar fractions with greater than ~$1.6R$/mol configuration entropy, where $R$ is the gas constant [3-5]. It is argued that a large configuration entropy (the so-called 'high-entropy' effect) can help to stabilize the solid solution phase against the formation of intermetallic compounds, and HEAs may also exhibit severe lattice distortion, sluggish diffusion, and 'cocktail' effects, but some of these core effects are in debate [3, 6]. HEAs exhibit a few general traits, such as low stacking fault energies, high thermal stability, good corrosion resistance, and improved radiation tolerance [3-5, 7-9]. It is also easier to design trade-offs of different properties in HEAs due to the availability of compositional spaces. While most early work was focused on single-phase HEAs, more recent studies have begun to investigate dual-phase and multi-phase HEAs [3-5, 10-16]. Notably, dual-phase (FCC + BCC or HCP) metallic HEAs have been explored extensively due to their superior and tunable mechanical properties [3-5, 10-14]. In this study, we explore dual-phase high-entropy ceramics (DPHECs) to further extend the families of compositionally-complex ceramics [17].

On the one hand, researchers have reported the fabrication of numerous high-entropy ceramics in bulk form in the last four years, including rocksalt [18], perovskite [19], and fluorite [20] oxides, borides [21], carbides [22-25], silicides [26, 27] and nitrides [28] (while high-entropy nitride and carbide thin films and coatings were also reported previously [29-31]). Most recently, Luo and co-workers propose to further broaden high-entropy ceramics to compositionally-complex ceramics (CCCs) [17, 32, 33]. See a recent review and perspective for related discussion [17].

In 2016, Gild et al. [21] reported the synthesis of $(Ti_{0.2}Zr_{0.2}Nb_{0.2}Hf_{0.2}Ta_{0.2})B_2$ and five other single-phase high-entropy borides (HEBs) of hexagonal $AlB_2$ metal diboride structure, representing the first high-entropy ultrahigh temperature ceramics (UHTCs) made in bulk form. Since then, HEBs [21, 34-36] and high-entropy carbides (HECs) [22-25, 37] such as $(Ti_{0.2}Zr_{0.2}Nb_{0.2}Hf_{0.2}Ta_{0.2})C$ with the rocksalt (B1) structure have been fabricated and studied extensively. In a recent study, HEB and HEC were fabricated (separately) via a reactive flash spark plasma sintering (reaFSPS) from mixtures of five commercial powders [34]; notably, single-phase HEB formed in 2 mins in reaFSPS (via "flash sintering" with a large current flowing through the specimen) and ~99% density was achieved with minimal oxides after using minor carbon additive [34]. This motivated us to adopt a small amount of carbon additive as a reducing agent and sintering aid to achieve high densities, but via a different (new) processing route, to fabricate a new class of dual-phase high-entropy UHTCs (DPHE-UHTCs) in this study.

In addition, researchers have been working on UHTCs to search for materials that can withstand extreme environments for potential applications in leading edges for hypersonic vehicles, nuclear reactors, armors, etc. [38-41]. Within the family of UHTCs, transition metal diborides (*e.g.*, $ZrB_2$, $HfB_2$, and $TaB_2$) and carbides (*e.g.*, ZrC, HfC, and TaC) are the most promising materials owing to their high melting temperatures and other outstanding properties [25, 42, 43]. Yet, none of these single-phase metal diborides and carbides can satisfy the whole gamut of demanding requirements [43, 44]. Moreover, dual-phase (or multi-phase) UHTCs are often used to achieve desired mechanical and oxidation resistant properties. Some common examples of this are $ZrB_2$-based or $HfB_2$-based UHTCs with secondary SiC, WC, and/or $B_4C$ phase(s) to achieve superior mechanical properties and oxidation resistance [44-47]. In addition, boride-carbide dual-phase UHTCs, e.g., $TiB_2$-TiC, $ZrB_2$-ZrC, and $NbB_2$-NbC, have been made by two-step sintering [48], spark plasma sintering (SPS) [49], and carbothermal reduction [50], which exhibit high melting/eutectic temperatures, good electronic conductivities, high flexural strength and hardness, and good wear resistance [51-53]. Dual-phase UHTCs also allow more tunable properties and more room for microstructural engineering. Yet, dual-phase high-entropy UHTCs (DPHE-UHTCs) have not been investigated to date.

In this work, we have fabricated the first HEB-HEC DPHE-UHTCs. Moreover, we have devised a novel synthesis and processing route to use $N$ individual borides and $(5 - N)$ carbides (each with different metals) as the start powders, along with 1 wt. % carbon additive as the sintering aid and reducing agent, to form two high-entropy phases (*i.e.*, HEB and HEC in a chemical equilibrium each other) via "reactive" SPS and achieved >98.5% of the theoretical densities (with the lowest measured relative density being ~98.8%) with virtually no native oxide contamination. This study further discovered a thermodynamic relation that governs the compositions of the HEB and HEC phases in equilibrium in DPHE-UHTCs. This new class of dual-phase high-entropy UHTCs exhibit several interesting properties, tunable by phase fraction. In a broader context, this study suggests dual-phase high-entropy ceramics or DPHECs as a new platform to tailor various properties via changing the phase fraction and microstructure. In general, the metal cation percentages in the two high-entropy phases should be different in a given DPHEC, where a thermodynamic equilibrium between the two high-entropy solution phases will likely be achieved and govern the (generally non-equal) partition of each metal element.

## 2. Experimental

### 2.1. Synthesis and Sintering

Commercial powders of $TiB_2$, TiC, $ZrB_2$, ZrC, $NbB_2$, NbC, $HfB_2$, HfC, TaC (99.5% purity, ~325 mesh, purchased from Alfa Aesar, MA, USA), and $TaB_2$ (99% purity, ~45 μm, purchased from Goodfellow, PA, USA) were used as the start powders for synthesizing the dual-phase high-entropy UHTCs. For each specimen listed in Table 1, appropriate amounts of five powders ($N$ boride powders and ($5 - N$) carbide powders as shown in Table 1, where $N = 0, 1, 2, 3, 4$ and 5, with the stoichiometry being calculated on the metal basis), were weighted out in batches of 10 g with 0.1 g (or ~1wt. %) of graphite (99.9% purity, 0.4-1.2 μm, purchased from US Nano, TX, USA) being added to the systems. The powders were hand mixed, and subsequently high energy ball milled (HEBM) in a Spex 8000D mill (SpexCertPrep, NJ, USA) in tungsten carbide lined stainless steel jars and 11.2 mm tungsten carbide milling media, at weight ratio between powder and milling media ~1:2.3, for 100 min with 0.1 g stearic acid as lubricant. The HEBM was performed in an argon atmosphere ($O_2 < 10$ ppm) with 50-min milling segments and 10-min cool-off periods to prevent overheating and oxidation.

The milled powders were loaded into 20-mm graphite dies lined with graphite foils in batches of 8 g, and subsequently consolidated into dense pellets via spark plasma sintering (SPS) in vacuum ($10^{-2}$ Torr) using a Thermal Technologies 3000 series SPS (Thermal Technologies, CA, USA). During SPS, the powders were first held at 1400 °C, and then at 1600 °C, respectively, for 80 min each to allow out-gassing as well as reduction of native oxides with the carbon additive, with minimal uniaxial load of 5 MPa at a heating rate of 100 °C /min. After that, the temperature was raised to 2200 °C at a slower rate of 30 °C /min and held at 2200 °C isothermally for 20 min for densification; at the same time, the uniaxial load was increased to and held at 80 MPa on a rate of 5 MPa/min.

After sintering, all specimens were cooled in the SPS machine to room temperature within 10-15 min. The final sintered pellets were measured to approximately 3-4 mm in thickness. After grinding and polishing the surfaces, the densities were measured via the Archimedes method with an accuracy of ± 0.01 g/cm$^3$.

Specimens of six different compositions were fabricated in this study. The specific conditions, including the starting powders and measured compositions for each specimen, are shown in Table 1. The specimens are called HEB, 8B2C, 6B4C, 4B6C, 2B8C and HEC, respectively, in this article, to represent the nominal HEB:HEC molar ratios of 100%-0%, 80%-20%, 60%-40%, 40%-60%, 20%-80% and 0%-100%, respectively. However, these are only the targeted nominal ratios and the actual measured compositions are also given in Table 1. The differences, when present, stem from the non-equal partition of metals in carbides and borides, as well as the introduction of a small amount of W from WC ball mill media.

## 2.2. Characterization

X-ray diffraction (XRD) was conducted on a Rigaku Miniflex diffractometer with Cu Kα radiation at 30 kV and 15 mA over a 2θ range of 20°-80° using 0.02° steps.

Scanning electron microscopy (SEM), electron dispersive X-ray spectroscopy (EDS), and electron backscatter diffraction (EBSD) analyses were conducted on a FEI Apreo microscope equipped with an Oxford N-Max$^N$ EDX detector and an Oxford Symmetry EBSD detector at an acceleration voltage of 5 kV, 30 kV, and 20 kV, respectively.

## 2.3. Measurements of Hardness and Moduli

Microhardness measurements were performed on all specimens with a Vickers diamond indenter at specified loading force of 1 kgf (9.8 N) or 200 gf (1.96 N) with a hold time of 15 seconds following the ASTM standard C1327. The size of the indentations was within the range of 25-30 and 10-15 μm for loading forces 9.8 and 1.96 N, respectively. Multiple measurements were performed at different locations on each specimen.

Moduli measurements were conducted with a Tektronix TDS 420A digital oscilloscope at 20 MHz for a longitudinal ultrasonic wave, and at 5 MHz for a transverse ultrasonic wave, following the ASTM standard E494-15. The longitudinal wave and transverse wave had average velocities in the range of 7000-9000 m/s and 4000-6000 m/s, respectively, in all specimens. Multiple measurements were carried out at different locations on each specimen to calculate the means and standard deviations.

## 2.4. Thermal Conductivity Measurements

Thermal conductivities were determined using the optical pump-probe technique steady-state thermoreflectance (SSTR), of which details can be found in Ref. [54]. A 532 nm continuous wavelength (CW) pump laser was used to induce steady-state temperature rise in the sample. A probe beam from 786 nm CW diode probe laser was used to detect the resulting reflectance change. SSTR uses Fourier's law to determine thermal conductivity by changing the pump power and monitoring the corresponding temperature rise [55, 56]. Before measurements, a thin Al layer (87 ± 4 nm, determined by picosecond acoustics [57]) was deposited on the sample surface by electron beam evaporation to act as an optical transducer. The pump and probe laser $1/e^2$ diameters were nearly equal, about 20 μm. SSTR measurements are nearly insensitive to transducer properties and heat capacity of the samples [54].

SSTR measurements require accurate determination of a proportionality constant $\gamma$, that relates the temperature change predicted via the thermal modeling to the measured change in surface reflectivity; thus, $\gamma$ is related to the thermoreflectance coefficient and conversion factor between reflectance change and photodetector voltage change. $\gamma$ is determined from a calibration sample and used in thermal conductivity measurements under the assumption that it remains constant between the sample and calibration. Sapphire was used as the calibration and the resultant $\gamma$ value was used to determine the thermal conductivity of Si and $z$-cut quartz; all values were found to be in good agreement with literature [54, 58, 59]. This $\gamma$ value was then used to determine the thermal conductivity of single-phase HEB and HEC, and dual-phase high-entropy UHTCs. As the surface roughness of the samples were slightly higher compared with the sapphire calibration, the uncertainty in determination of $\Upsilon$ value was ~10%.

## 3. Results

### 3.1. Dual-phase Microstructure

SEM micrographs from backscattered electrons for all sintered specimens are illustrated in Fig. 1. All HEB-HEC DPHE-UHTCs demonstrate distinctive dual-phase microstructures in Fig. 1(b)-1(e), where the lower-density HEB phases show a darker contrast while the higher-density HEC phases exhibit a brighter contrast. HEB and HEC phases are indicated by arrows in Fig. 2(a) at a higher magnification. Not surprisingly, the area fraction of the dark HEB phase decreases, while that of the HEC phase increases, monotonously from 8B2C to 2B8C.

A small amount (<1-1.5 vol.%) of very dark spots can be observed in Fig. 1, which were identified as either pores or remaining graphite additive. Both have similar (very dark) contrasts at low magnifications. They are only distinguishable at high magnifications as shown in Fig. 2(b), where the surface morphology/features inside the pores can be observed by SEM.

To further quantify the volumetric ratios of the HEB and HEC phases in these DPHE-UHTCs, digital image processing was conducted on low-magnification (100×) SEM micrographs for all dual-phase specimens; specific detail of this procedure is given in Supplementary Fig. S1. The measured volumetric percentages of HEB and HEC phases are plotted in Fig. 3 and listed in Table 1. Each volumetric percentage was computed from multiple SEM micrographs at different locations of each specimen; the measured values were averaged and rounded to the closest integer. The measured HEB (and HEC) vol.% and mol.% are normalized to the total HEB + HEC amount (excluding a total amount of <1.5 vol.% of the pores and remaining graphite that is infeasible to quantify exactly). The notation of 8B2C, 6B4C, 4B6C, and 2B8C are the nominal compositions of the specimens; figures are all plotted based on the measured mol.% of the HEC (carbide) phase.

### 3.2. Formation of High-Entropy Boride and Carbide Phases

XRD analyses were carried out for all specimens to determine the HEB and HEC phase structures and confirm the phase purity. Fig. 4 shows the XRD patterns of all specimens after HEBM and after SPS, respectively. After HEBM, multiple distinct hexagonal and cubic phases are detected (Fig. 4(a)), which shows that the high-entropy solid solution phases did not yet form before the SPS. The XRD peak broadening observed in Fig. 4(a) can be attributed to particle and grain size reduction and mechanical alloying from the HEBM. After SPS, all observed peaks can be attributed to one hexagonal $AlB_2$ phase and one cubic rocksalt phase, which corresponds to the HEB phase and HEC phase, respectively. The variations in the relative XRD intensities from 8B2C to 2B8C are consistent with changes in the HEB vs. HEC phase fractions, which are evident in, and quantified based on, SEM micrographs. The lattice parameters for each individual phase were calculated and are listed in Table 1.

Specimen HEC with five carbide precursors shows a single rocksalt phase in its XRD pattern. Specimen HEB with five boride precursors shows a dominant single hexagonal $AlB_2$ phase, albeit a small amount of secondary carbide phase, which presumably formed from the addition of 1 wt. % graphite (to reduce/remove native oxides and promote sintering). This secondary carbide phase in Specimen HEB can also be seen in the SEM micrograph, which are the bright spots in Fig. 1(a). A point EDS analysis (Supplementary Fig. S6) showed that it is a high-entropy carbide or HEC phase akin to those found in dual-phase 8B2C to 2B8C specimens (instead of WC debris particles from ball mill media). It appears that this minor carbide (HEC) phase (of only ~1.5 vol. %) is in a chemical equilibrium with the primary HEB phase, and its measured composition (Supplementary Fig. S6) follows the same thermodynamic relationship for the HEB-HEC DPHE-UHTCs discussed subsequently in §4.2.

### 3.3. Compositions of High-Entropy Phases

EDS elemental analysis was conducted separately on each of the two phases. The metal atomic percentages of Ti, Zr, Nb, Hf, and Ta in the HEB and HEC phases were measured, and the results are illustrated in Fig. 5(a) and 5(b), respectively. In the HEB phase, the Ti composition increases monotonously from ~25% in 8B2C to 40% in 2B8C, the Zr and Nb compositions remain roughly constant at ~19%, and Hf (and Ta) compositions decrease continuously from 18% to 12% (and from 15% to 7%), with the increasing HEC fraction. Here, all percentages are at. % on the metal basis (excluding B or C), unless otherwise noted. In the HEC phase, the Ti composition increases monotonously from 10% to 17%, the Zr and Nb compositions fluctuate at ~16%, and the Hf (and Ta) compositions decreases from 25% to 21% (and from 33% to 24%), with the increasing HEC fraction (from 8B2C to 2B8C). On the other hand, all five cations (Ti, Zr, Nb, Hf, and Ta) have a similar composition of ~18-20% as expected in single-phase HEB and HEC specimens. Here, means and standard deviations were calculated from multiple measurements at different locations on the specimens. Because of the peak overlapping of Zr and Nb, Hf and Ta in EDS spectra, atomic percentages of these cations possess larger uncertainties.

Furthermore, 1-6% of W, from the WC milling jar and media, is present in both the HEB and HEC phases of all specimens. The W compositions are also plotted accordingly in Fig. 5(a) and 5(b). The measured compositions of both HEC and HEB phases for all specimens are summarized in Table 1.

### 3.4. Compositional Homogeneity

To verify the compositional homogeneity of the HEB and HEC phases, EDS elemental mapping was conducted at a micrometer-scale for all specimens and the results are shown in Fig. 6. In the single-phase HEB and HEC specimens, all elements are uniformly distributed. Four of the DPHE-UHTCs, Fig. 6(b)-(e) show different concentrations in the individual HEB and HEC phases. Consistent with the quantitative compositional analysis discussed above, Ti, Zr and Nb are enriched in the HEB phases, while Hf and Ta are enriched in the HEC phases. Notably, the composition is highly uniform within each of the two (HEB and HEC) phases, which suggests the formation of homogeneous high-entropy boride and carbide solid solutions. The W maps were also collected but are not shown here, because the signals are barely above the background noises due to the low concentrations.

### 3.5. Densities

Based on the compositions measured from EDS and the lattice parameters obtained from XRD, theoretical density for each HEB or HEC phase was calculated. With the HEB and HEC volumetric percentage hitherto attained by digital image processing, the theoretical densities of all specimens were further determined. Using the experimental densities measured via Archimedes method, the relative densities of all sintered specimens were determined to be between ~98.8% and 100%. Specifically, Specimen 2B8C has the lowest measured relative density of ~98.8%, and all other specimens are >99% dense. This observation is consistent with results from digital image processing, where the total combination of pores and graphite regions with dark contrasts were measured to be less than 1.5 vol. % in all cases. The measured actual and relative densities for all specimens are given in Table 1.

### 3.6. Grain Size and Microstructure

EBSD was utilized to measure the grain size distribution and examine the possible texture of all sintered specimens. For the DPHE-UHTCs, EBSD was conducted independently on the HEB and HEC phases. The EBSD maps and grain size distributions for each phase are shown in Fig. 7. As shown in Fig. 7(a1), 7(b1), 7(c1), and 7(d1), the average grain size for the boride (HEB) phase decreases from 8.2 μm in 8B2C to 4.2 μm in 2B8C with decreasing HEB phase fraction. At the same time, the average grain size of carbide (HEC) phase increases from 4.9 μm to 11.4 μm with the increasing HEC phase fraction, as shown in Fig. 7(a2), 7(b2), 7(c2), and 7(d2).

The EBSD maps and grain size distributions for the single-phase HEB and HEC specimens are displayed in Fig. 8. The averaged grain size of the single-phase HEB specimen was measured to be 15.0 μm, while that of single-phase HEC specimen was measured to be 12.1 μm. Both are greater than the averaged grain sizes of their counterparts in DPHE-UHTCs.

At the same time, EBSD maps revealed no significant texture for any of the specimens. The inverse pole figures of crystal preferred orientation for dual-phase (8B2C, 6B4C, 4B6C, and 2B8C) and single-phase (HEB and HEC) UHTCs can be found in Fig. S2 and Fig. S3, respectively, in Supplementary Material.

The grains of both the HEB and HEC phases are largely equiaxed for all six specimens. The grain aspect ratios on each of the boride and carbide phases for all these UHTCs are displayed in Fig. S4 in Supplementary Material.

### 3.7. Vickers Microhardness

Fig. 9 illustrates the measured Vickers microhardness for the sintered specimens at indent loading force of 9.8 N. The Vickers microhardness of the single-phase HEB was measured to be $19.4 \pm 1.3$ GPa, and that of the HEC was determined to be $25.3 \pm 1.4$ GPa. All DPHE-UHTCs have the measured Vickers hardness values in between these single-phase specimens. The hardness increases substantially from single-phase HEB to 8B2C, and subsequently increases somewhat linearly, but more moderately, with the increasing HEC phase fraction. Meanwhile, measured Vickers microhardness at a different indent loading force of 1.96 N is illustrated in Supplementary Fig. S5. A similar trend of microhardness change from HEB to HEC is also observed at this loading force with slightly higher (by 0.8-1.5 GPa) measured hardness values for all specimens.

### 3.8. Moduli

Young's modulus of elasticity ($E$) and shear modulus ($G$) for all sintered specimens were measured and are shown in Fig. 10(a) and 10(b), respectively. The measured Young's modulus decreases monotonously from $524.6 \pm 6.9$ GPa for the single-phase HEB to $462.4 \pm 4.0$ GPa for the single-phase HEC specimen. The measured moduli for DPHE-UHTCs are in between those of HEB and HEC. The measured shear modulus also decreases monotonously from $232.2 \pm 4.6$ GPa for the single-phase HEB to $193.0 \pm 3.6$ GPa for the single-phase HEC specimen. The measured shear moduli for DPHE-UHTCs follows a more linear relation.

### 3.9. Thermal Conductivity

Fig. 11 displays the measured thermal conductivity of all specimens. The thermal conductivity decreases from ~26 Wm$^{-1}$K$^{-1}$ (for HEB, 8B2C, and 6B4C) to ~17 Wm$^{-1}$K$^{-1}$ (for 4B6C and 2B8C), and then further to ~13 Wm$^{-1}$K$^{-1}$ (for HEC); it generally follows a trend of monotonous decrease (with rather large error bars) with most substantial (step-wise) reduction observed between 6B4C and 4B6C.

### 4. Discussion

### 4.1. A New and Novel Processing Route

In the current processing, each metal was only present in either boride or carbide form (but not both) in the initial mixture of five powders of binary carbides or borides (see Table 1 for the specific start mixtures of powders for each case). Subsequently, the cations will be partitioned to both the HEB and HEC phases after SPS in the final specimens, presumably in a chemical equilibrium with one another. Thus, this process may be considered as a "reactive SPS" since it involves a chemical reaction, albeit it is not a reactive sintering directly from metal and B/C elements. We should note that sintering methods involving alternative chemical reactions (e.g. boro-carbothermal reduction of metal oxides) may also be regarded as "reactive sintering" [60, 61].

In addition to adding minor graphite, a strategy to minimize the native oxides is to adopt carbides of the metals that are more prone to oxidations, *e.g.*, ZrC and HfC (instead of $ZrB_2$ and $HfB_2$ that typically have more native oxides in the initial commercial powders), as the starting powders, whenever possible.

A key achievement of the current work is the new and novel processing route to attain great than ~98.8% relative densities with virtually no native oxides and minimum impurity phase. This has been achieved via utilizing 1 wt. % graphite as the reducing agent and sintering aids, along with further optimization of processing (e.g., incorporating a two-step prolonged pre-heating stages at 1400°C and 1600°C) to allow reduction and out-gassing. On top of that, holding the specimen at 2200°C for 20 mins at 80 MPa also facilitates the densification process.

The total amount of porosity plus impurity phases (mostly remaining graphite and possibly $B_4C$) in the final sintered specimens is less than 1.5 vol. % for all specimens, as confirmed independently by digital image processing. This novel processing itself is an advancement from the state of the art of sintering high-entropy UHTCs, particularly in comparison with prior studies of fabricating HEBs from commercial powders [21]. In the past, high-density HEBs had to be made by synthesized high-entropy powders [62, 63] or via special method of reaFSPS [34].

It has been widely accepted that existence of native oxides would hinder densification in both boride and carbide systems [48, 64, 65]. In our study, 1 wt. % graphite was added as an in situ reducing agent and powder processing was conducted in an argon atmosphere. Extra carbon in the specimens can help to keep a local reducing environment during both HEBM and SPS. Moreover, pellets were held at 1400°C and 1600°C for longer periods of time before the final densification. Annealing at 1400°C in vacuum promotes removal of intrinsic metal oxides and $B_2O_3$ with the assistance of extra graphite [48]. Holding at 1600°C, together with a low ramping rate of 30 °C /min during densification, can facilitate pore elimination before rapid grain growth [66]. The combination of these strategies in this new procedure helped us to achieve the high relative densities (with the lowest measured relative density of ~98.8%, and >99% in all but this one cases) with virtually no native oxides observed in the final specimens. This has not been achieved previously for HEBs made from commercial powders via normal HEBM and SPS.

We note that W contamination from WC balling media is an inevitable issue with the current processing route. In the current experiments, the overall measured W percentages are consistent across all specimens in the range of 3.0% to 5.0%. In general, the levels of W contamination can depend on the processing conditions and change as the WC media progressively worn. Thus, careful measurements of the actual compositions in the final specimens should be conducted to calibrate the nominal compositions, and the analysis should be based on the actual measured/calibrated (instead of nominal) compositions.

### 4.2. High-Entropy Boride-Carbide DPHE-UHTCs

Combining the HEB and HEC phase percentages measured from the digital image processing, the lattice parameters measured from XRD, and compositions of each phase measured from EDS, the molar fractions of HEB and HEC phases were also calculated. The actual molar phase percentages and compositions were determined for all specimens, as:

- HEB: 98% $(Ti_{0.22}Zr_{0.19}Nb_{0.18}Hf_{0.19}Ta_{0.19}W_{0.03})B_2$ + a minor carbide phase
- 8B2C: 76% $(Ti_{0.25}Zr_{0.19}Nb_{0.20}Hf_{0.18}Ta_{0.15}W_{0.03})B_2$ + 24% $(Ti_{0.10}Zr_{0.13}Nb_{0.14}Hf_{0.25}Ta_{0.33}W_{0.05})C$;
- 6B4C: 55% $(Ti_{0.30}Zr_{0.21}Nb_{0.19}Hf_{0.16}Ta_{0.10}W_{0.04})B_2$ + 45% $(Ti_{0.12}Zr_{0.16}Nb_{0.16}Hf_{0.23}Ta_{0.29}W_{0.04})C$;
- 4B6C: 35% $(Ti_{0.35}Zr_{0.19}Nb_{0.20}Hf_{0.15}Ta_{0.08}W_{0.03})B_2$ + 65% $(Ti_{0.13}Zr_{0.17}Nb_{0.18}Hf_{0.22}Ta_{0.24}W_{0.06})C$;
- 2B8C: 17% $(Ti_{0.40}Zr_{0.20}Nb_{0.20}Hf_{0.12}Ta_{0.07}W_{0.01})B_2$ + 83% $(Ti_{0.17}Zr_{0.17}Nb_{0.17}Hf_{0.21}Ta_{0.24}W_{0.04})C$; and
- HEC: $(Ti_{0.20}Zr_{0.21}Nb_{0.21}Hf_{0.18}Ta_{0.17}W_{0.03})C$, respectively.

The results above are summarized and tabulated in Table 1. Noticeably, all high-entropy dual-phase UHTC specimens contain higher molar percentages (3-5%) of the carbide (HEC) phase than the nominal fractions. These discrepancies stem from several causes: (1) the addition 1 wt. % of graphite initially as a reducing agent and sintering aid, (2) the contamination from WC milling jar and media, and (3) the evaporation of $B_2O_3$ [48] during sintering. For the same reason, a minor carbide phase of ~1.5 vol. % is observed in the (nominally) HEB specimen. As we have discussed earlier in §3.2, this minor carbide phase is also a HEC phase in an chemical equilibrium with the primary HEB phase, with its composition being governed by the same thermodynamic relationship (Supplementary Fig. S6). The mol. % of the boride phase is estimated from the measured volumetric percentage and the cation ratios in HEB vs. HEC phase derived in the next section (since it is difficult to accurately measure the composition of the minor carbide phase); the error should be negligibly small (well below the 1% round-off error), given the small volumetric fraction of the secondary carbide phase. It should also be pointed out that the carbon vacancies commonly observed in transition metal carbides [67] (also noted as $MC_{1-x}$, where $M$ is the metal atom) shouldn't be significant in all HEC phases above due to excess carbon in the systems from extra 1 wt.% graphite addition.

The non-equimolar partitions of the metal cations between the HEC and HEB phases also results in the deviation of the nominal and actual phase fractions. As shown in Fig. 5(a) and 5(b), the phase compositions of the HEB and HEC phases can vary significantly in the four dual-phase high-entropy UHTCs, and they depend on the phase fraction, although there are roughly equal molar amounts of each metal cations overall (*i.e.*, 20% each to start with, albeit extra W contamination picked up from HEBM).

### 4.3. Equilibrium Compositions of the HEB and HEC Phases

The ratios of cation/metal percentages in HEB *vs.* HEC phases were calculated and plotted against the actual molar fraction of the HEC phase in Fig. 12. Although the cation compositions vary significantly in the HEB and HEC phases with the varying phase fraction, this ratio remains remarkably steady. First, Ti preferentially dissolves into the HEB phase with the ratio of $Ti_{HEB}$:$Ti_{HEC} \approx 2.5$:1. Second, both Zr and Nb exhibit slight preference in dissolving in the HEB phase, with the ratios of $Zr_{HEB}$:$Zr_{HEC} \approx 1.3$:1 and $Nb_{HEB}$:$Nb_{HEC} \approx 1.2$:1, respectively. Third, both Ta and Hf are enriched in the HEC phase, with ratios of $Hf_{HEB}$:$Hf_{HEC} \approx 1$:1.5 and $Ta_{HEB}$:$Ta_{HEC} \approx 1$:2.9, respectively. The nearly constant ratios, largely independent of the HEC/HEB phase fraction, also support that chemical equilibria were likely achieved in this set of specimens. In summary, the following relation approximately holds for the HEB and HEC phases in chemical equilibria at 2200°C:

$$\begin{cases} \frac{X_{Ti}^{HEB}}{X_{Ti}^{HEC}} \approx 2.5 & (\sim 0.25 < X_{Ti}^{HEB} < \sim 0.40; \sim 0.10 < X_{Ti}^{HEC} < 0.17) \\ \frac{X_{Zr}^{HEB}}{X_{Zr}^{HEC}} \approx 1.3 & (\sim 0.19 < X_{Zr}^{HEB} < \sim 0.21; \sim 0.13 < X_{Zr}^{HEC} < 0.17) \\ \frac{X_{Nb}^{HEB}}{X_{Nb}^{HEC}} \approx 1.2 & (\sim 0.17 < X_{Nb}^{HEB} < \sim 0.20; \sim 0.14 < X_{Nb}^{HEC} < 0.18) \\ \frac{X_{Hf}^{HEB}}{X_{Hf}^{HEC}} \approx 0.67 & (\sim 0.12 < X_{Hf}^{HEB} < \sim 0.18; \sim 0.21 < X_{Hf}^{HEC} < 0.25) \\ \frac{X_{Ta}^{HEB}}{X_{Ta}^{HEC}} \approx 0.36 & (\sim 0.07 < X_{Ta}^{HEB} < \sim 0.15; \sim 0.24 < X_{Ta}^{HEC} < 0.33) \end{cases}, \quad (1)$$

where $X_M^{HEB}$ and $X_M^{HEC}$ ($M$ = Ti, Zr, Nb, Hf or Ta) are the fraction (or percentage) of the $M$ in the HEB and HEC phases, respectively, on the metal basis.

The preferential dissolution of metal cations in HEB *vs.* HEC phases can be justified from the formation energies for different transition metal diborides and carbides calculated from density functional theory (DFT) [68]. By comparing the formation energies of metal diboride ($E_f^{MB_2}$) and carbide ($E_f^{MC}$) for a same metal cation [69], it can be found that the differential formation energies ($E_f^{MB_2} - E_f^{MC}$) are $-0.25$ eV/atom for Ti, $-0.24$ eV/atom for Nb, $-0.18$ eV/atom for Zr, $-0.08$ eV/atom for Hf, and $-0.07$ eV/atom for Ta, respectively. The order is largely the same as that in $X_M^{HEB}/X_M^{HEC}$ with the only exception of the order of Zr and Nb. Note that the formation energy from elements of the metal diboride is always lower than that of the carbide of the same metal (so that $E_f^{MB_2} - E_f^{MC} < 0$ for all cases), but only the relative rankings are relevant for considering the preferential dissolution in HEB vs. HEC with the overall mass conservation and stoichiometry requirements imposed on the dual-phase equilibria in closed systems.

A simplified ideal solution model can be proposed to further rationalize the observed composition ratios in HEB *vs.* HEC phases. When an HEB phase and an HEC phase are in a thermodynamic equilibrium, the chemical potential for each metal $M$ cation in the HEB and HEC should be equal ($\mu_M^{HEB} = \mu_M^{HEC}$). For ideal solid solutions (as the first order of approximation), we have:

$$\mu_M^{HEB} = \mu_M^{MB_2} + RTlnX_M^{HEB} \tag{2}$$

and

$$\mu_M^{HEC} = \mu_M^{MC} + RTlnX_M^{HEC}, \tag{3}$$

where $\mu_M^{MB_2}$ and $\mu_M^{MC}$ are the chemical potentials for the metal $M$ cation in their corresponding individual diboride and carbide, respectively. At a given temperature, we have

$$\mu_M^{MB_2} - \mu_M^{MC} = \left(\Delta\overline{H}_M^{MB_2} - \Delta\overline{H}_M^{MC}\right) - T\left(\Delta\overline{S}_M^{MB_2} - \Delta\overline{S}_M^{MC}\right) \tag{4}$$

where $\Delta\overline{H}$ and $\Delta\overline{S}$ are the corresponding molar partial enthalpy and entropy of $M$, respectively. In the above equation, the molar partial entropy is the vibration entropy $\Delta\overline{S}_{vib}$, as the configuration entropy $\Delta\overline{S}_{config}$ in the single metal diboride or carbide is zero. Since the vibrational entropy $\Delta\overline{S}_{vib}$ depends mainly on the crystal structure [70], $T\left(\Delta\overline{S}_{M,vib}^{MB_2} - \Delta\overline{S}_{M,vib}^{MC}\right)$ should be relatively independent of the specific metal cation at a fixed temperature. Moreover, the differential molar partial enthalpy of $M$ in pure $M$B$_2$ and $M$C should be correlated with the differential formation energies $\left(E_f^{MB_2} - E_f^{MC}\right)$ calculated from DFT as:

$$\left(\Delta\overline{H}_M^{MB_2} - \Delta\overline{H}_M^{MC}\right) = \left(E_f^{MB_2} - E_f^{MC}\right) + C_1 \tag{5}$$

To the first order of approximation, combining Eqs. (2-5) produces:

$$RTln\frac{X_M^{HEB}}{X_M^{HEC}} = -\left(E_f^{MB_2} - E_f^{MC}\right) + C_2, \tag{6}$$

where $C_2$ is a largely constant (as the first order of approximation). Eq. (6) shows that $X_M^{HEB}/X_M^{HEC}$ ratios are expected from the ideal solution models, which is consistent with our experimental observations (Fig. 12 and Eq. (1)). To further analyze our experimental results quantitively, a linear regression of $RTln\left(X_M^{HEB}/X_M^{HEC}\right)$ from our experimentally measured $\left(X_M^{HEB}/X_M^{HEC}\right)$ values (as given in Eq. (1)) *vs.* DFT calculated differential formation energies $\left(E_f^{MB_2} - E_f^{MC}\right)$ values was performed, which produces:

$$RTln\frac{X_M^{HEB}}{X_M^{HEC}} = -1.65\left(E_f^{MB_2} - E_f^{MC}\right) - 26303 \text{ J/mol} \tag{7}$$

with the correlation coefficient:

$$r = 0.903 \text{ (or } r^2 = 0.816) \tag{8}$$

The errors in the $\left(E_f^{MB_2} - E_f^{MC}\right)$ values calculated from DFT [71, 72], as well as the two simplified assumptions of ideal solutions and constant $\left(\Delta\bar{S}_{M,vib}^{MB_2} - \Delta\bar{S}_{M,vib}^{MC}\right)$ (i.e., independent of the specific metal $M$), should have contributed to the nonlinearity ($r < 1$) and the deviation of the fitted slope from the theoretical value of $-1$ in Eq. (6). Nonetheless, the agreement with experiments is rather satisfactory given all the errors, simplifications, and assumptions.

Overall, the proposed thermodynamic model supports the constant $X_M^{HEB}/X_M^{HEC}$ ratios observed in experiments, and it largely produces the correct orders of the experimentally observed $X_M^{HEB}/X_M^{HEC}$ ratios of five metal cations (with only one exception of swamping the positions of Zr and Nb).

## 4.4. Microstructural Development

There is a definite correlation between the grain sizes of the HEB and HEC phases with the phase fraction, as shown in Fig. 13. With very limited mutual solubilities of the anions between the HEB and HEC phases, the two phases inhibit the grain growth of each other [73, 74]. This is illustrated by the measured averaged grain sizes in Fig. 13 (from the EBSD maps in Fig. 7 and Fig. 8), where both boride and carbide phases show significant larger grain size in single-phase UHTCs (HEB and HEC), in comparison with those in DPHE-UHTCs (8B2C, 6B4C, 4B6C, and 2B8C). Among the four DPHE-UHTCs, the carbide phase in 8B2C and the boride phase in 2B8C have the smallest averaged grain sizes (4.9 μm and 4.2 μm, respectively). This can be explained by their low molar fractions (24% and 17%, accordingly), as it is more difficult for the minor phase to grow directly. As observed in the EBSD maps in Fig. 7(a2) and Fig. 7(d1), the grains in all these phases are mostly isolated by their complementary phase in the same specimen, which are then prevented from further grain growth. Moreover, the (mean-field) coarsening is small because of the limited mutual solubilities of anions.

## 4.5. Hardness

The measured Vickers hardness and Young's moduli are summarized in Table 2. As we can see from previous studies of HEB [63] and HEC [22, 37], Vickers hardness demonstrates a trend of indentation load dependence, *viz.*, the measured Vickers hardness value decreases when the indentation load increases. For HEB, our measured Vickers hardness is consistent with that reported by Zhang *et al.* [62] (19.4 ± 1.3 vs. 21.7 ± 1.1 GPa) despite different loading forces (9.8 N vs. 1.96 N); and at the same loading force of 1.96 N, the discrepancy becomes even smaller (20.2 ± 1.3 vs. 21.7 ± 1.1 GPa). In the first work of HEB, Gild *et al.* [21] reported a lower Vickers hardness value of 17.5 ± 1.2 GPa, which can be attributed to the lower relative density. Gu *et al.* [63] reported the change in measured Vickers hardness from ~22GPa to ~26GPa when the indent loading force decreased from 9.8 N to 1.96 N, whereas our measurement indicates hardness of 19.4 and 20.2 GPa at loading forces of 9.8 and 1.96 N respectively; this discrepancy may be ascribed to (1) different synthesis method (boro/carbothermal reduction by Gu *et al.* [63]), and (2) the associated residual phases in the specimens (harder $B_4C$ in boro/carbothermal reduction and softer C in the current case).

For HEC, the measured nanohardness values (obtained from nanoindentation) are significantly greater than the measured microhardness values (30-40 vs. 15-30 GPa). Our measured Vickers hardness (25.3 ± 1.4 GPa at 9.8 N and 26.8 ± 1.6 GPa at 1.96 N) is similar to that presented by Ye *et al.* [37] (18.8 ± 0.4 GPa at 9.8 N and 22.5 ± 0.3 GPa at 0.98 N). In addition, a lower value of 15 GPa at 9.8 N indent loading force was reported in Ref. [23]. Besides the different indent loading forces applied in different studies, the relatively low Vickers hardness can also be attributed to their low relative densities (95.3% and 93%, respectively, in comparison with the high relative density of ~99% in this study).

It has already been shown by several independent studies that the single-phase HEB [21, 37, 62] or HEC [22, 24] is harder than the rule-of-mixture (RoM) average of those of the five individual components (binary carbides or borides). A new observation in this study is that the Vickers hardness of a dual-phase UHTC is further enhanced above the linear interpolation of two single-phase endmembers (HEB and HEC), which is indicated by the blue dash line in Fig. 9; on the other hand, hardness data taken at indent loading force of 1.96 N in Fig. S5 also confirms this phenomenon. A commentary here is that the grains in DPHE-UHTCs (4.2-11.9 μm in size) are substantially smaller than those of single-phase HEB and HEC counterparts (with size of 15.0 μm and 12.1 μm, respectively); with higher hardness being reported on finer grain ceramics [75, 76],the actual mechanism of this enhancement on Vickers hardness for DPHE-UHTCs would require further study.

### 4.6. Moduli

Young's and shear moduli of the DPHE-UHTCs are further assessed. Here, the theoretical RoM averages from the data in Material Project Database [69] are presented in Fig. 10 as the references. Here, the rule of mixture has been applied twice. First, it was used to obtain the weighted averaged moduli for HEB and HEC phases, respectively, based on the actual measured composition of each phase. Second, the RoM averages of the dual-phase high-entropy UHTCs were obtained from weighted averages of the moduli of two single-phase HEB and HEC calculated in Step 1.

One may argue that traditional isostrain model of Voigot and isostress model of Reuss for dual-phase materials adopted in RoM are overly simplified. In a more generalized approximation recently developed by Zhang et al. [77], Young's and shear moduli for dual-phase material can be estimated as $E = E_1 \frac{E_2(f_2 + \delta f_1) + E_1(f_1 - \delta f_1)}{E_2 \delta f_1 + E_1(1 - \delta f_1)}$ and $G_1 \frac{G_2(f_2 + \delta f_1) + G_1(f_1 - \delta f_1)}{G_2 \delta f_1 + G_1(1 - \delta f_1)}$, where $E_1$, $G_1$, $f_1$ and $E_2$, $G_2$, $f_2$ are Young's modulus, shear modulus, and volume fractions of the primary and secondary phases, respectively; and $\delta$ is the shape and distribution parameter, which is set to be 1/3 for three-dimensional cubic systems. We also applied the above equations to our dual-phase high-entropy UHTCs, and the results show negligible discrepancies with those obtained from simple RoM averages, shown in Supplementary Fig. S7.

Elastic moduli of two single-phase UHTCs (HEB and HEC) are tabulated in Table 2 along with those reported in previous publications. Fig. 10 shows that the general trends of experimental and calculated values conform each other. However, there are several interesting and new observations, which will be discussed subsequently.

A comparison of selected moduli of the HEB and HEC phases with the reported values show our measurements are consistent with literature. Gu et al. [63] used a mechanical resonance frequency technique to determine Young's moduli, which is similar to our method. They have measured the Young's modulus of their high relative density (97.9%) HEB specimen (with the same nominal composition as our HEB) to be 527 GPa, which is within the uncertainty of our measurement (524.6 ± 6.9 GPa). At the same time, they also ascribed the low Young's modulus (500 GPa) to more porosity in their low relative density (94.4%) HEB specimen. For HEC, on the other hand, the Young's modulus measured in this study (462.4 ± 4.0 GPa) is close to those reported by Yan et al. (479 GPa) [23] and Harrington et al. (443 ± 40 GPa) [22], although both previous studies applied nanoindentation in the measurement. Ye et al. [37] reported the Young's modulus of HEC to be 514 - 522 ± 10 GPa, where the disparity might be attributed to the small indent loading force (8 mN) in their nanoindentation measurement [22]. Shear moduli for single-phase HEB and HEC UHTCs are rather limited in previous studies. Sarker et al. [24] reported the shear modulus of HEC as 188 GPa, which is consistent with our measurement of 193.0 ± 3.6 GPa in this study. In summary, our measured moduli are consistent with those reported in literature for single-phase high-entropy UHTCs, whenever there are reported data, albeit with expected differences due to porosity and loading.

For the four DPHE-UHTCs (8B2C, 6B4C, 4B6C, and 2B8C), both their elastic and shear moduli follow a largely linear interpolation with their single-phase endmembers (HEB and HEC); this trend is also consistent with that expected from a composite rule.

The measured Young's moduli for the single-phase HEB and HEC specimens, as well as the dual-phase high-entropy UHTCs, are higher than the theoretical RoM averages from individual components in all cases. Similar observations have already been reported for single-phase HEB and HEC previously [22-24, 63], yet our Young's moduli measured from ASTM standard acoustic wave tests are expected to be more accurate than those obtained via nanoindentation.

A most interesting new observation in Fig. 10(b) is that the measured shear modulus of HEC is higher than the theoretical RoM value, while the opposite is observed for HEB. We assume this phenomenon is related to different bonding nature of the layered $AlB_2$ structure [78]. As observed in Table 1 and Ref. [21], interlayer lattice parameters $c$ measured experimentally from XRD patterns are always larger than those predicted by the RoM of individual components; this can be understood intuitively, as large cations will weighted more in determining the lattice parameter $c$ in layered structure with rigid covalent B layer that separates the metal layers, leaving more space around smaller cations. This expansion along the $c$-axis implies a relatively weak interlayer metal-B bonding, which would make the structure more susceptible to shear deformation. The (~0.3%) expansion of the lattice parameters $c$ of the HEB (2.359 Å) with respective to RoM average (3.348 Å) has been confirmed by XRD for the current case/composition (Table 1), and this effect (expanded $c$) was observed universally for HEBs of many different compositions [21].

It should be noted that porosity can contribute to the decrease of measured moduli for all six UHTCs above. Nevertheless, based on the empirical relationship proposed by Dean et al. [79], the porosity in these UHTCs can only affect their measured moduli by <2-3% due to their high relative density (≥99%, except for 2B8C ≈98.8%); this impact is smaller than or barely comparable to the uncertainties associated with the measured moduli hence excluded in the discussion above.

### 4.7. Thermal Conductivity

The uncertainty in the thermal conductivity value incorporates the standard deviation, the uncertainty in determining $Υ$ value, transducer thermal conductivity, and the thermal boundary conductance between Al transducer and sample.

The thermal conductivities of the single-phase HEB and HEC are significantly lower than those of their constituent borides and carbides. HEB has a thermal conductivity of $26.2 \pm 2.8$ $Wm^{-1}K^{-1}$, whereas $ZrB_2$ can achieve a thermal conductivity of over 100 $Wm^{-1}K^{-1}$ [80-84], though the value may vary widely based on processing conditions, porosity, impurity concentrations, and microstructure. The thermal conductivity of HEC is $13.2 \pm 1.7$ $Wm^{-1}K^{-1}$, whereas TiC and TaC have thermal conductivities of $30.4 \pm 1.3$ and $36.2 \pm 1.5$ $Wm^{-1}K^{-1}$, respectively [85]. Such decrease in thermal conductivity agrees with reported thermal conductivity reduction in high entropy silicide [26] and entropy stabilized rocksalt oxide systems [86, 87].

For the dual-phase specimens, the most significant drop in thermal conductivity, observed between 6B4C and 4B6C, can be attributed to the change of specimen matrix phase [88, 89], as the carbide phase ratio passes 50 vol. %. As shown in Fig. 1 and 7, HEB phase performs as the matrix in 8B2C and 6B4C, whereas 4B6C and 2B8C have HEC phase as their matrix, largely due to the different phase volume ratios in each specimen.

### 4.8. Tunable Properties

In comparison with their single-phase counterparts, this new class of dual-phase high-entropy UHTCs provide a new platform to tune properties, including grain size, hardness, modulus, and thermal conductivity that have already been demonstrated here as well as many other functional properties. Furthermore, DPHE-UHTCs also provide more possibilities for controlling mechanical and other properties via microstructural engineering, *e.g.*, thru changing the fraction, the size, and the shape of each phases, as well as tailoring the interfaces. These should be explored systematically in future studies.

In a broader context, we expect that various physical properties can be tuned and enhanced in dual-phase high-entropy ceramics (DPHECs), just like metallic dual-phase HEAs with demonstrated unique and tunable mechanical properties [3-5, 90-93].

## 5. Conclusions

This study successfully fabricates, for the first time to our knowledge, a new series of dual-phase high-entropy UHTCs, with high-entropy boride and carbide solid-solution phases formed in equilibria with each other, where a thermodynamic relation exists to govern the non-equal partition of each metal elements.

In addition to systematically exploring dual-phase high-entropy ceramics (DPHECs) for the first time, this study also presents the following new observations or advancements:

- A new reactive SPS route is developed to make single-phase and dual-phase high-entropy UHTCs directly from commercial powders to achieve ~99% or higher relative densities with virtual no native oxides via utilizing 1 wt. % graphite as a reducing agent and sintering aid, which represent a significant advancement in the processing of high-entropy UHTCs (particularly high-entropy borides that have been proven difficult to densify). Noticeably, 3-5% W contamination from HEBM is observed in all sintered specimens.

- We discovered a thermodynamic relation that dictates the compositions of the HEB and HEC phases in equilibrium and drives them away from equimolar composition (despite the nearly equimolar composition overall for the DPHE-UHTCs), which is important and essential to design dual-phase high-entropy UHTCs. In general, the metal cation fractions in the two high-entropy phases in any dual-phase high-entropy ceramics (DPHEC) should be different; the two high-entropy solid-solution phases are likely in an equilibrium with each other, where a thermodynamic relation exists and governs the (generally non-equal) partition of each of the metal elements.

- The grain sizes and properties of dual-phase high-entropy UHTCs can be tailored by changing phase fraction and microstructure.

- The hardness of the dual-phase high-entropy UHTCs are improved from the weighted linear average of the two single-phase high-entropy UHTCs, which are already harder than the rule-of mixture averages of individual binary carbides and borides.

- The Young's moduli of high-entropy UHTCs, which are measured accurately from acoustic waves, are higher than the theoretical rule-of mixture predictions.

- As a new observation, the shear modulus of HEC is higher, but that of the HEB is lower, than the rule-of mixture average; the latter unusual observation is explained from different crystal structures and a confirmed expansion of interlayer distance in the layered HEB structures.

- The measured thermal conductivities of single-phase HEB and HEC are significantly lower than those of their constituent borides and carbides. A stepwise transition of thermal conductivity in the DPHE-UHTCs is revealed. The relatively large decrease between 6B4C and 4B6C can be ascribed to the change of boride matrix to carbide matrix.

In a broader content, the current study extends the state of the art by introducing dual-phase high-entropy ceramics or DPHECs to provide a new platform to tailor and further enhance various functional properties.

**Acknowledgements**

This work is primarily supported by an Office of Naval Research MURI program (grant no. N00014-15-1-2863). We thank our program managers, Dr. Kenny Lipkowitz and Dr. Eric Wuchina, and all MURI colleagues for guidance, encouragement, and helpful scientific discussion. We also acknowledge Dr. Lei Chen, who had conducted some earlier exploration of fabricating HEB-HEC dual-phase composites via a different processing recipe.

# Supplementary Information

Supplementary Table S1 and Supplementary Figures S1-S7 related to this article can be found, in the online version, at doi: xxxxxxxx.

**Table 1.** Summary of the six compositions studied. Specimens HEB and HEC are single-phase, while 8B2C, 6B4C, 4B6C, and 2B8C are (HEB + HEC) DPHE-UHTCs. Compositions of all HEB and HEC phases were measured from EDS. The volume percentages of HEB and HEC phases were measured from digital imaging processing. Note that the carbide (HEC) vol. % and mol. % in Table 1 are normalized to the total HEC + HEB amount (excluding <1-1.5 vol. % of the pores and remaining graphite in total). The experimental lattice parameters were measured by XRD, whereas averaged values represent the weighted means of single metal borides or carbides calculated via the rule of mixture. See Supplementary Table S1 for additional data.

| Specimen | Precursors | Post-sintering Boride and Carbide Phase Compositions | | Experimental Lattice Parameters a, c (Å) | Averaged Lattice Parameters by RoM a, c (Å) | Theoretical Phase Density (g/cm³) | HEB-HEC (vol. %) | HEB-HEC (mol. %) | Theoretical Density (g/cm³) | Measured Density (g/cm³) (Relative Density) |
|---|---|---|---|---|---|---|---|---|---|---|
| HEB | $20TiB_2$-$20ZrB_2$-$20NbB_2$-$20HfB_2$-$20TaB_2$ | Boride | $(Ti_{0.22}Zr_{0.19}Nb_{0.18}Hf_{0.19}Ta_{0.19}W_{0.03})B_2$ | 3.097, 3.359 | 3.105, 3.348 | 8.34 | 98.5% | 98% | 8.34 | 8.27 (≈99%) |
| | | Carbide | - | - | - | - | - | - | | |
| 8B2C | $20TiB_2$-$20ZrB_2$-$20NbB_2$-$20HfC$-$20TaB_2$ | Boride | $(Ti_{0.25}Zr_{0.19}Nb_{0.20}Hf_{0.18}Ta_{0.15}W_{0.03})B_2$ | 3.103, 3.367 | 3.103, 3.346 | 7.99 | 79% | 76% | 8.67 | 8.62 (>99%) |
| | | Carbide | $(Ti_{0.10}Zr_{0.13}Nb_{0.14}Hf_{0.35}Ta_{0.33}W_{0.05})C$ | 4.514 | 4.516 | 11.23 | 21% | 24% | | |
| 6B4C | $20TiB_2$-$20ZrC$-$20NbB_2$-$20HfC$-$20TaB_2$ | Boride | $(Ti_{0.30}Zr_{0.15}Nb_{0.19}Hf_{0.16}Ta_{0.16}W_{0.04})B_2$ | 3.098, 3.360 | 3.100, 3.347 | 7.49 | 59% | 55% | 8.80 | 8.80 (≈100%) |
| | | Carbide | $(Ti_{0.12}Zr_{0.16}Nb_{0.16}Hf_{0.33}Ta_{0.19}W_{0.04})C$ | 4.513 | 4.514 | 10.69 | 41% | 45% | | |
| 4B6C | $20TiC$-$20ZrC$-$20NbB_2$-$20HfC$-$20TaB_2$ | Boride | $(Ti_{0.35}Zr_{0.20}Nb_{0.20}Hf_{0.17}Ta_{0.08}W_{0.08})B_2$ | 3.095 3.355 | 3.095, 3.337 | 7.23 | 39% | 35% | 9.20 | 9.12 (≈99%) |
| | | Carbide | $(Ti_{0.15}Zr_{0.21}Nb_{0.18}Hf_{0.22}Ta_{0.24}W_{0.06})C$ | 4.508 | 4.513 | 10.46 | 61% | 65% | | |
| 2B8C | $20TiC$-$20ZrC$-$20NbB_2$-$20HfC$-$20TaC$ | Boride | $(Ti_{0.40}Zr_{0.25}Nb_{0.20}Hf_{0.13}Ta_{0.07}W_{0.07})B_2$ | 3.089, 3.346 | 3.092, 3.331 | 6.85 | 20% | 17% | 9.37 | 9.26 (≈98.8%) |
| | | Carbide | $(Ti_{0.17}Zr_{0.07}Nb_{0.07}Hf_{0.07}Ta_{0.24}W_{0.04})C$ | 4.515 | 4.508 | 10.01 | 80% | 83% | | |
| HEC | $20TiC$-$20ZrC$-$20NbC$-$20HfC$-$20TaC$ | Boride | - | - | - | - | - | - | 9.34 | 9.30 (>99%) |
| | | Carbide | $(Ti_{0.20}Zr_{0.21}Nb_{0.21}Hf_{0.18}Ta_{0.17}W_{0.03})C$ | 4.505 | 4.509 | 9.34 | 100% | 100% | | |

**Table 2.** Comparison of Vickers hardness and Young's modulus of HEB and HEC measured in this study and those reported in literature. The indent loading forces applied in the hardness measurements, which affect the reported hardness values, are also given.

| Composition | Relative Density | Measured Vickers Hardness (GPa) | Indent Loading Force | Young's Modulus (GPa) | Reference |
|---|---|---|---|---|---|
| HEB ($Ti_{0.2}Zr_{0.2}Nb_{0.2}Hf_{0.2}Ta_{0.2}$)$B_2$ | ≈99% | 19.4 ± 1.3 | 9.8 N | 524.6 ± 6.9 | This Work |
| | | 20.2 ± 1.0 | 1.96 N | | |
| | 94.4% | 22.44 ± 0.56 | 9.8 N | 500 | [63] |
| | | 25.61 ± 0.83 | 1.96 N | | |
| | 97.9% | 26.82 ± 1.77 | 1.96 N | 527 | [63] |
| | 96.3% | 21.7 ± 1.1 | 1.96 N | - | [62] |
| | 92.4% | 17.5 ± 1.2 | 1.96 N | - | [21] |
| HEC ($Ti_{0.2}Zr_{0.2}Nb_{0.2}Hf_{0.2}Ta_{0.2}$)C | >99% | 25.3 ± 1.4 | 9.8 N | 462.4 ± 4.0 | This Work |
| | | 26.8 ± 1.6 | 1.96 N | | |
| | 93% | 15 | 9.8 N | 479 | [23] |
| | 99% | 32 ± 2 | 300 mN | 443 ± 40 | [22, 24] |
| | 95.3% | 18.8 ± 0.4 | 9.8 N | 514 − 522 ± 10 | [37] |
| | | 22.5 ± 0.3 | 0.98 N | | |
| | | 40.6 ± 0.6 | 8 mN | | |

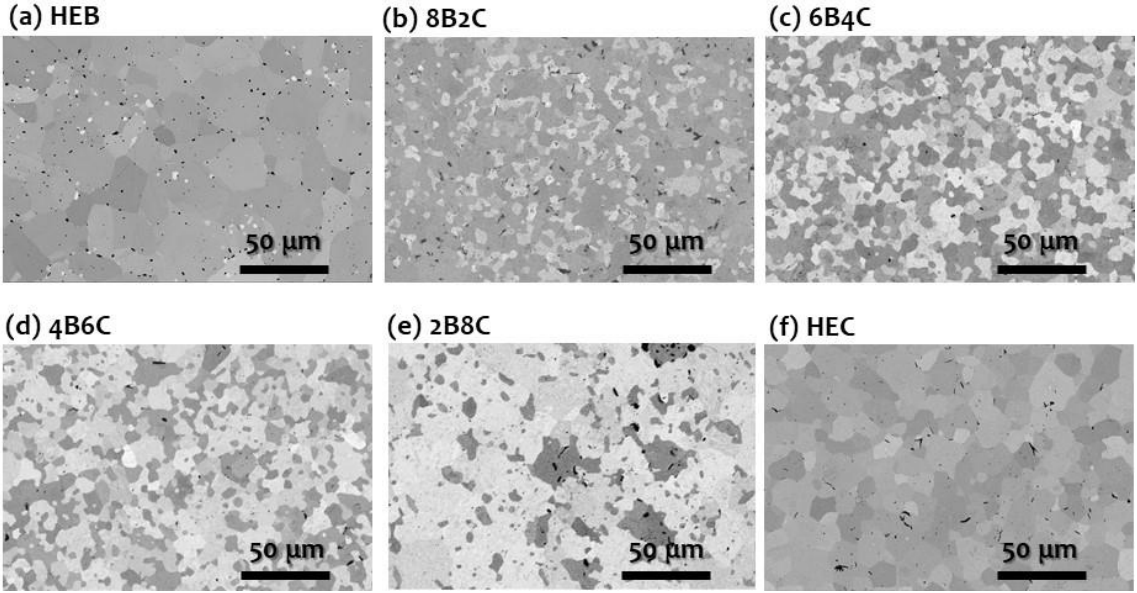

**Fig. 1.** SEM micrographs of sintered specimens: **(a)** HEB: 98% $(Ti_{0.22}Zr_{0.19}Nb_{0.18}Hf_{0.19}Ta_{0.19}W_{0.03})B_2$ + a minor carbide phase, **(b)** 8B2C: 76% $(Ti_{0.25}Zr_{0.19}Nb_{0.20}Hf_{0.18}Ta_{0.15}W_{0.03})B_2$ + 24% $(Ti_{0.10}Zr_{0.13}Nb_{0.14}Hf_{0.25}Ta_{0.33}W_{0.05})C$, **(c)** 6B4C: 55% $(Ti_{0.30}Zr_{0.21}Nb_{0.19}Hf_{0.16}Ta_{0.10}W_{0.04})B_2$ + 45% $(Ti_{0.12}Zr_{0.16}Nb_{0.16}Hf_{0.23}Ta_{0.29}W_{0.04})C$, **(d)** 4B6C: 35% $(Ti_{0.35}Zr_{0.19}Nb_{0.20}Hf_{0.15}Ta_{0.08}W_{0.03})B_2$ + 65% $(Ti_{0.13}Zr_{0.17}Nb_{0.18}Hf_{0.22}Ta_{0.24}W_{0.06})C$, **(d)** 4B6C: 35% $(Ti_{0.40}Zr_{0.20}Nb_{0.20}Hf_{0.12}Ta_{0.07}W_{0.01})B_2$ + 83% $(Ti_{0.17}Zr_{0.17}Nb_{0.17}Hf_{0.21}Ta_{0.24}W_{0.04})C$, and **(f)** HEC: $(Ti_{0.20}Zr_{0.21}Nb_{0.21}Hf_{0.18}Ta_{0.17}W_{0.03})C$. All phase percentages are calculated molar percentages based on XRD results and phase volume percentages from digital image processing.

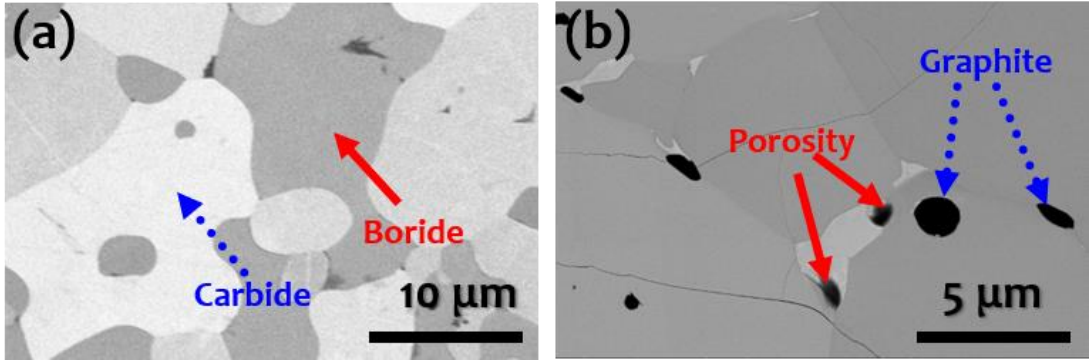

**Fig. 2.** SEM micrographs at higher magnifications. **(a)** Specimen 4B6C, where the boride (HEB) phase with a dark contrast and carbide (HEC) phase with a light contrast are indicated by the arrows. **(b)** Specimen HEB, where the porosity and remaining graphite are marked by arrows.

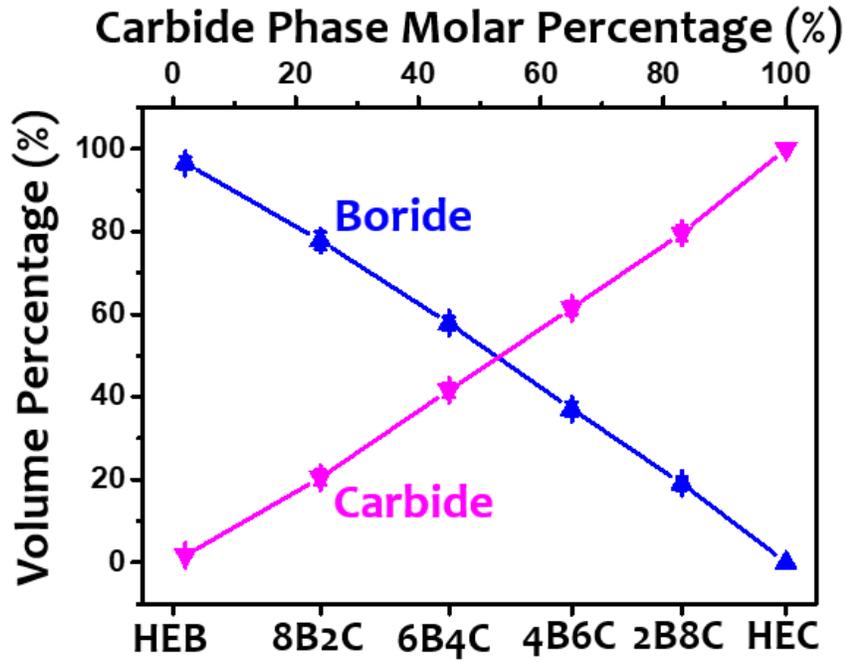

**Fig. 3.** The volume percentages of the HEB (boride) and HEC (carbide) phases, obtained via digital image processing, for all six UHTCs. The actual HEC (carbide) phase molar percentage is marked on the top *x*-axis. There is a minor carbide phase in the (nominally) HEB specimen due to graphite addition and carbon contamination from HEBM and SPS.

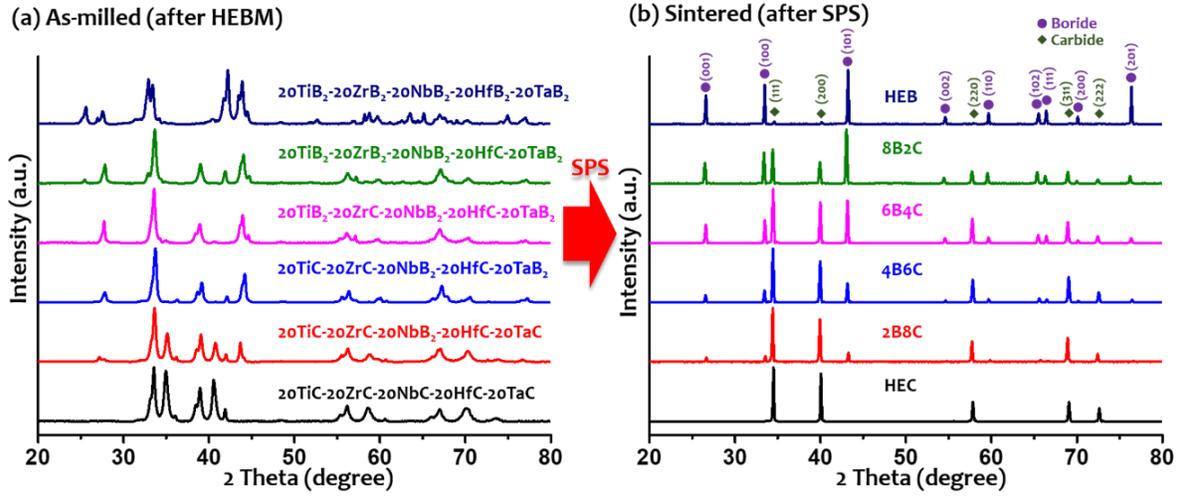

**Fig. 4.** XRD patterns of the samples of all six compositions (**a**) after HEBM (as-milled powders) and (**b**) after SPS (sintered pellets).

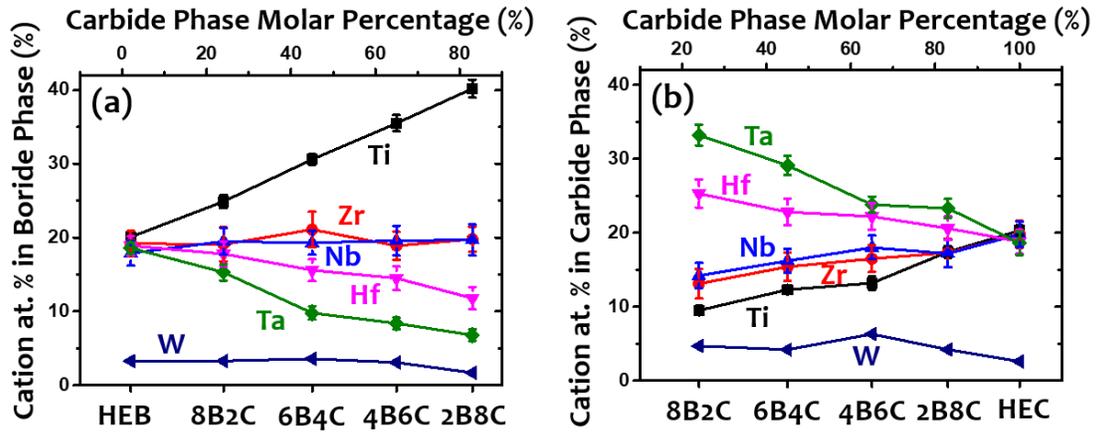

**Fig. 5.** Cation atomic percentages in (**a**) HEB (boride) phase and (**b**) HEC (carbide) phase, measured from EDS elemental analysis, for four DPHE-UHTCs (8B2C, 6B4C, 4B6C, and 2B8C), and HEB, HEC (nominally) single-phase UHTCs.

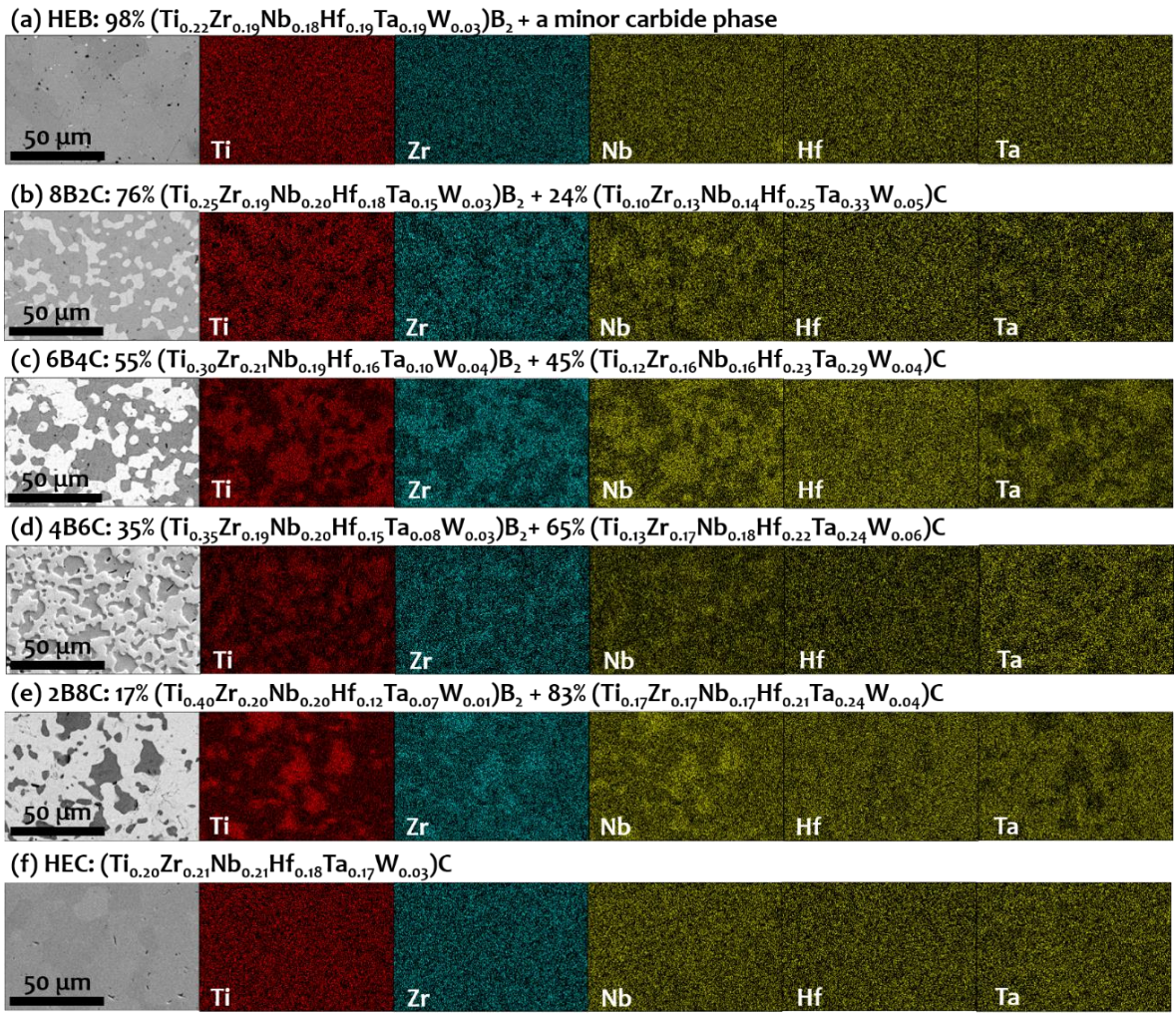

**(a) HEB:** 98% $(Ti_{0.22}Zr_{0.19}Nb_{0.18}Hf_{0.19}Ta_{0.19}W_{0.03})B_2$ + a minor carbide phase

**(b) 8B2C:** 76% $(Ti_{0.25}Zr_{0.19}Nb_{0.20}Hf_{0.18}Ta_{0.15}W_{0.03})B_2$ + 24% $(Ti_{0.10}Zr_{0.13}Nb_{0.14}Hf_{0.25}Ta_{0.33}W_{0.05})C$

**(c) 6B4C:** 55% $(Ti_{0.30}Zr_{0.21}Nb_{0.19}Hf_{0.16}Ta_{0.10}W_{0.04})B_2$ + 45% $(Ti_{0.12}Zr_{0.16}Nb_{0.16}Hf_{0.23}Ta_{0.29}W_{0.04})C$

**(d) 4B6C:** 35% $(Ti_{0.35}Zr_{0.19}Nb_{0.20}Hf_{0.15}Ta_{0.08}W_{0.03})B_2$ + 65% $(Ti_{0.13}Zr_{0.17}Nb_{0.18}Hf_{0.22}Ta_{0.24}W_{0.06})C$

**(e) 2B8C:** 17% $(Ti_{0.40}Zr_{0.20}Nb_{0.20}Hf_{0.12}Ta_{0.07}W_{0.01})B_2$ + 83% $(Ti_{0.17}Zr_{0.17}Nb_{0.17}Hf_{0.21}Ta_{0.24}W_{0.04})C$

**(f) HEC:** $(Ti_{0.20}Zr_{0.21}Nb_{0.21}Hf_{0.18}Ta_{0.17}W_{0.03})C$

**Fig. 6.** SEM micrographs and EDS elemental maps of Specimens **(a)** HEB, **(b)** 8B2C, **(c)** 6B4C, **(d)** 4B6C, **(e)** 2B8C, and **(f)** HEC. All samples were synthesized by the same procedure. Note that the boride (HEB) phases are enriched in Ti, Zr, and Nb, while the carbide (HEC) phases are enriched in Hf and Ta in the DPHE-UHTCs.

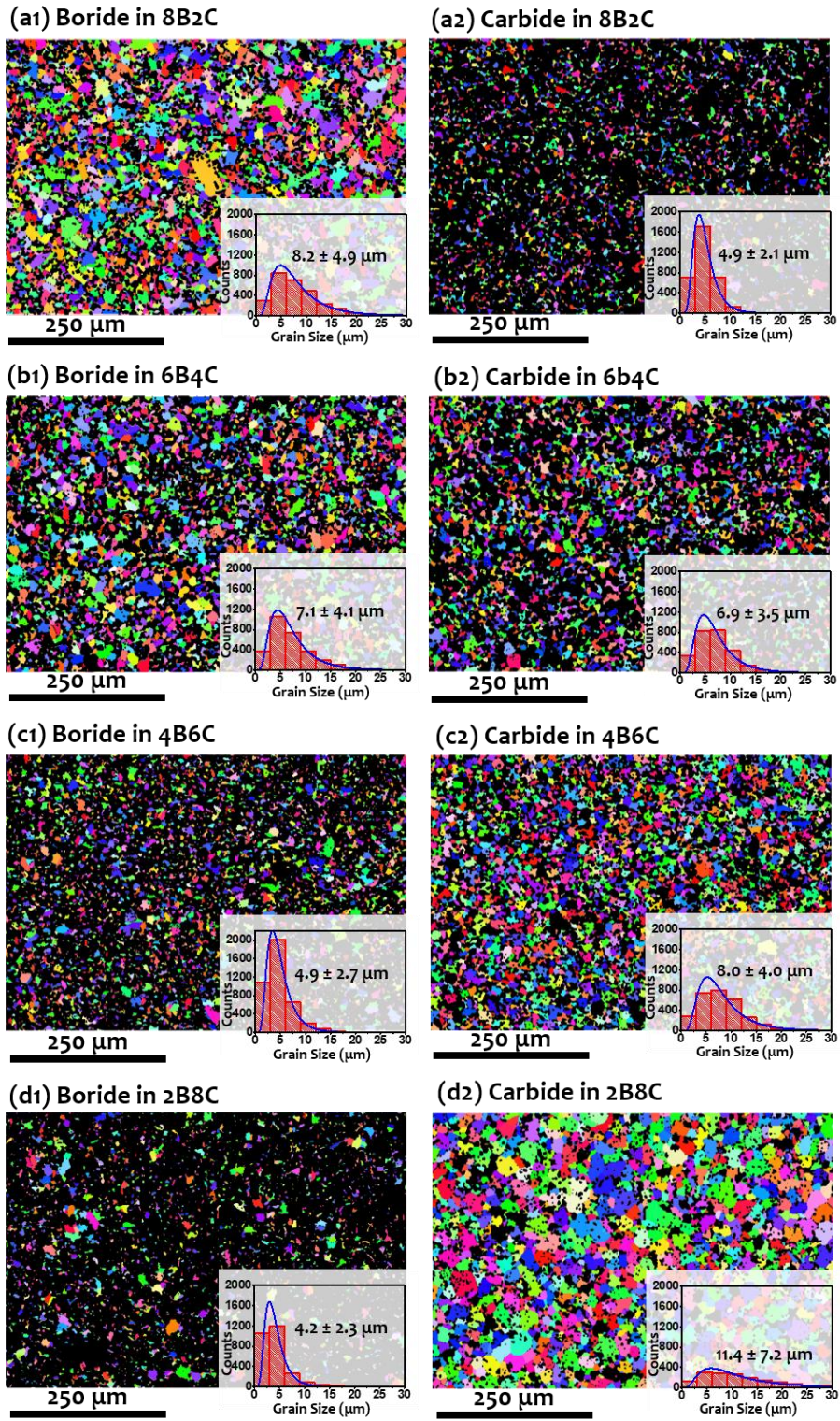

**Fig. 7.** EBSD maps and grain size distributions for Specimens **(a)** 8B2C, **(b)** 6B4C, **(c)** 4B6C, and **(d)** 2B8C. The EBSD maps of the boride (HEB) phases are displayed in the panels (a1), (b1), (c1), and (d1), while those of the carbide (HEC) phases are displayed in the panels (a2), (b2), (c2), and (d2). The insets are grain size distributions.

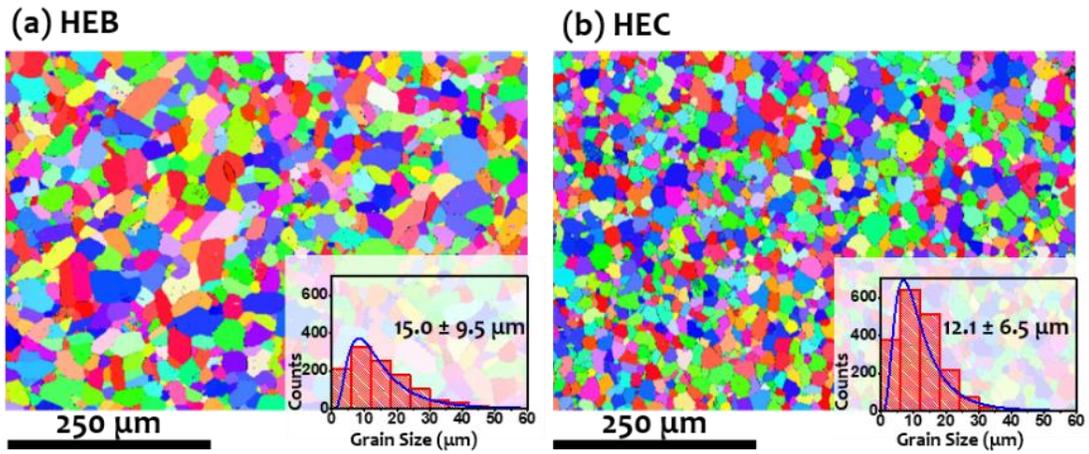

**Fig. 8.** EBSD maps and grain size distributions for Specimens **(a)** HEB and **(b)** HEC.

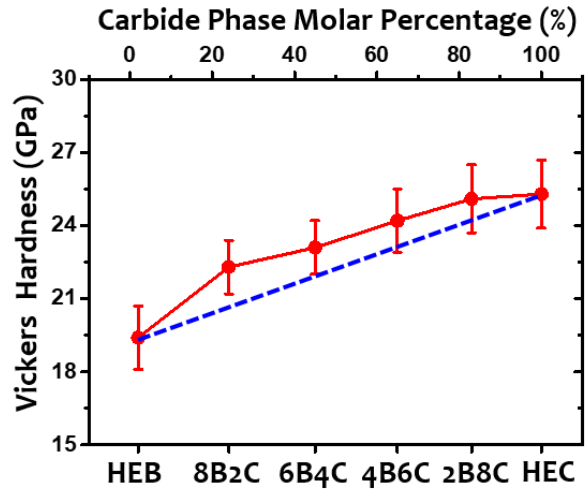

**Fig. 9.** Measured Vickers hardness values (red circles) of six specimens (HEB, 8B2C, 6B4C, 4B6C, 2B8C, and HEC) at indent loading force of 9.8 N. The blue dash line indicates the predicted hardness of DPHE-UHTCs from a simple linear composite rule.

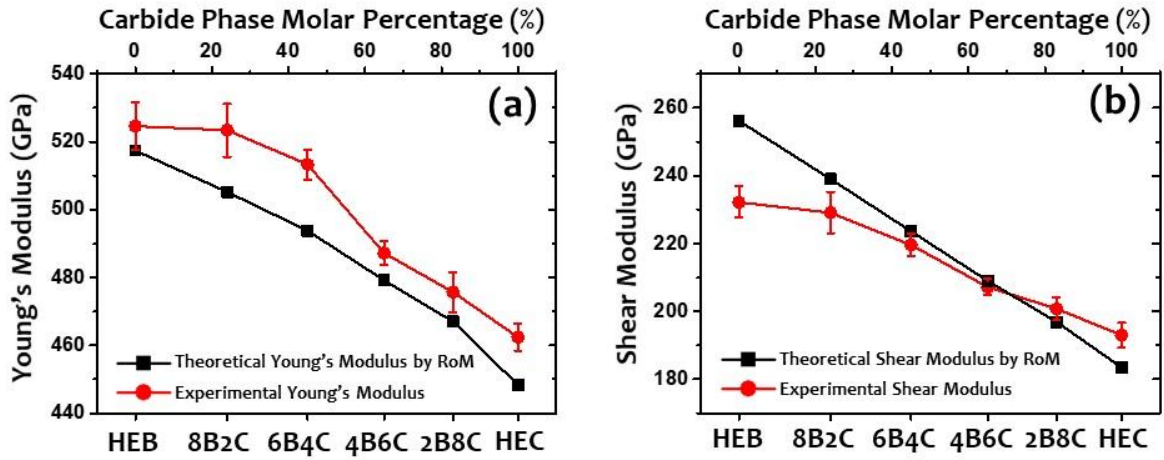

**Fig. 10.** **(a)** Young's moduli, and **(b)** shear moduli of six specimens (HEB, 8B2C, 6B4C, 4B6C, 2B8C, and HEC). The theoretical rule-of-mixture (RoM) average values are also presented as black squares as references.

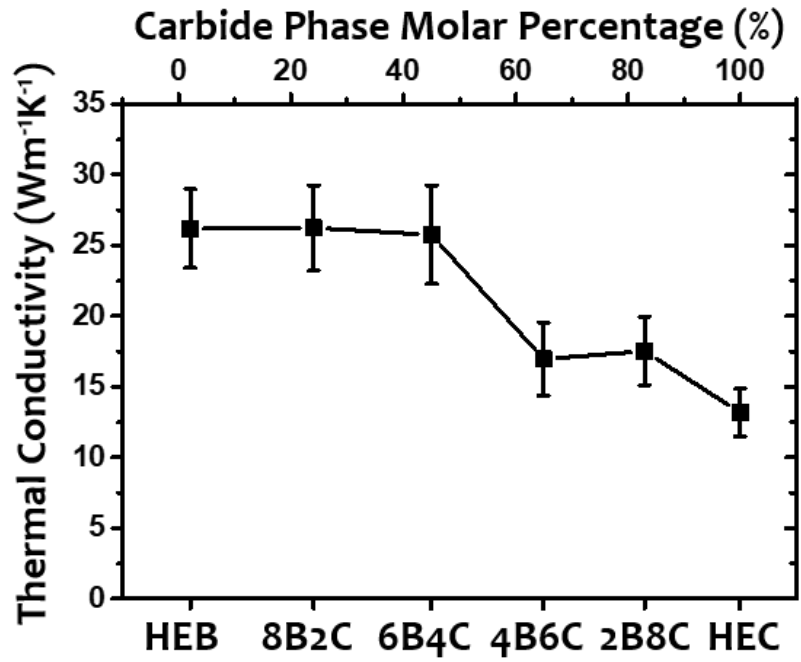

**Fig. 11.** Measured thermal conductivities of six specimens (HEB, 8B2C, 6B4C, 4B6C, 2B8C, and HEC).

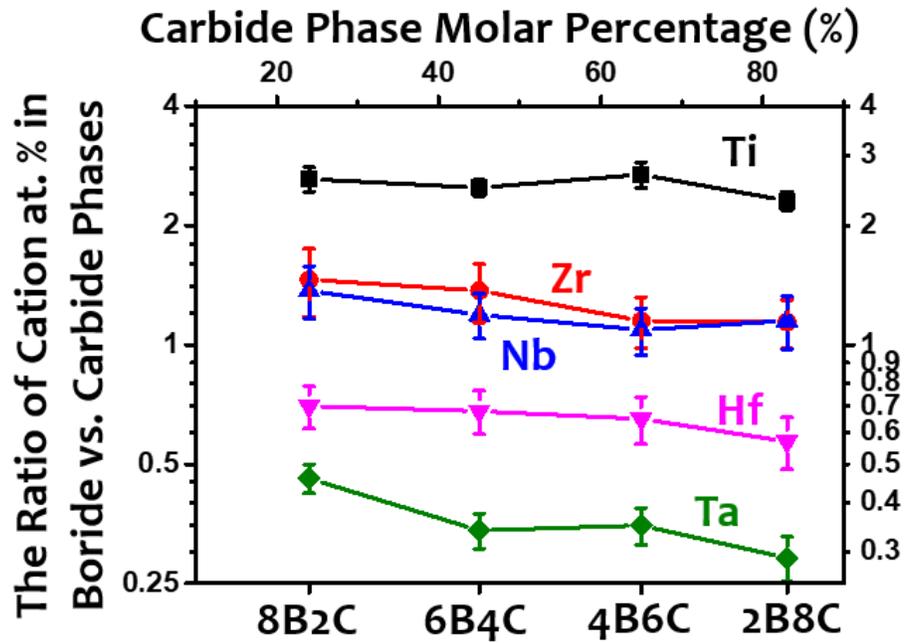

**Fig. 12.** The ratio of cation atomic percentages in HEB (boride) vs. HEC (carbide) phases, for four DPHE-UHTCs. Ticks with different values at the same scale are presented on the right y-axis.

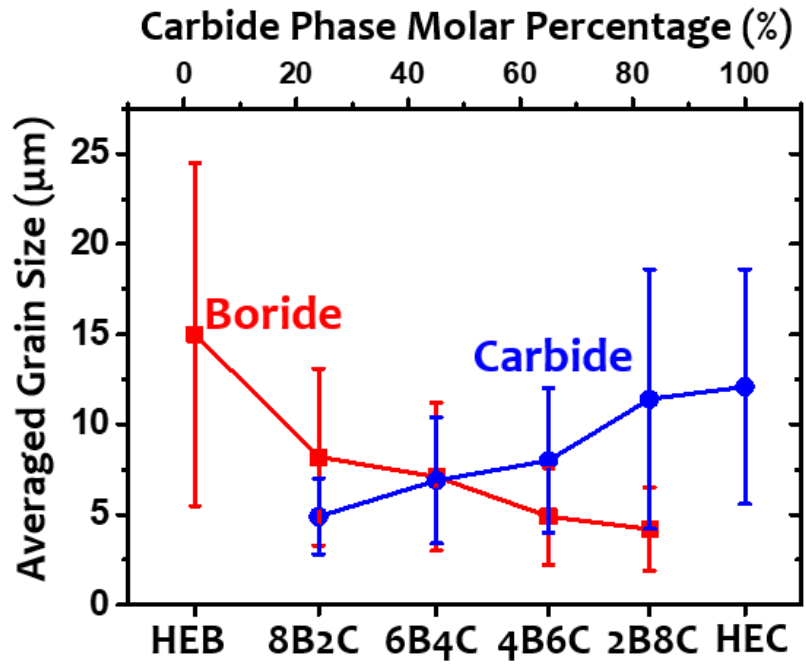

**Fig. 13.** The averaged grain size and distribution of the HEB (boride) and the HEC (carbide) phases, illustrated with red squares and blue dots respectively, for six specimens (HEB, 8B2C, 6B4C, 4B6C, 2B8C, and HEC).